\documentclass{aa}  
\usepackage{natbib,twoopt}
\usepackage[breaklinks=true]{hyperref} 
\bibpunct{(}{)}{;}{a}{}{,}             
\makeatletter
  \newcommandtwoopt{\citeads}[3][][]{\href{http://adsabs.harvard.edu/abs/#3}%
    {\def\hyper@linkstart##1##2{}%
     \let\hyper@linkend\@empty\citealp[#1][#2]{#3}}}
  \newcommandtwoopt{\citepads}[3][][]{\href{http://adsabs.harvard.edu/abs/#3}%
    {\def\hyper@linkstart##1##2{}%
     \let\hyper@linkend\@empty\citep[#1][#2]{#3}}}
  \newcommandtwoopt{\citetads}[3][][]{\href{http://adsabs.harvard.edu/abs/#3}%
    {\def\hyper@linkstart##1##2{}%
     \let\hyper@linkend\@empty\citet[#1][#2]{#3}}}
  \newcommandtwoopt{\citeyearads}[3][][]%
    {\href{http://adsabs.harvard.edu/abs/#3}
    {\def\hyper@linkstart##1##2{}%
     \let\hyper@linkend\@empty\citeyear[#1][#2]{#3}}}
\makeatother

\usepackage{graphicx}
\usepackage{txfonts}
\usepackage{subfigure}
\usepackage{longtable}
\usepackage{tabularx, blindtext}

\begin{document} 

 \title{Eight more low luminosity globular clusters in the Sagittarius dwarf galaxy}
   \author{
          D. Minniti\inst{1,2}
          \and
          M. Gómez\inst{1}
          \and
          J. Alonso-García\inst{3,4}
         \and
          R. K. Saito\inst{5}
          \and
          E. R. Garro\inst{1}
%
%
          }
\institute{Departamento de Ciencias Físicas, Facultad de Ciencias Exactas, Universidad Andrés Bello, Fernández Concha 700, Las Condes, Santiago, Chile
\and
 Vatican Observatory, Vatican City State, V-00120, Italy
\and
Centro de Astronomía (CITEVA), Universidad de Antofagasta, Av. Angamos 601, Antofagasta, Chile
\and
Millennium Institute of Astrophysics, Santiago, Chile
\and 
Departamento de Fisica, Universidade Federal de Santa Catarina, Trindade 88040-900, Florianopolis, SC, Brazil
}
  \date{Received; Accepted}

 
  \abstract
   {
The Sagittarius (Sgr) dwarf galaxy is merging with the Milky Way, and the study of its globular clusters (GCs) is important to understand the history and outcome of this ongoing process.
}
   {
   Our main goal is to characterize the GC system of the Sgr dwarf galaxy. This task is hampered by high foreground stellar contamination, mostly from the Galactic bulge.
   }
   {
     We performed a  GC search specifically tailored to find new GC members 
     within the main body of this dwarf galaxy
     using the combined data of the VISTA Variables in the Via Lactea Extended Survey (VVVX) near-infrared survey and the Gaia Early Data Release 3 (EDR3) optical database.
}
   {
We applied proper motion (PM) cuts to discard foreground bulge and disk stars, and we found a number of GC candidates in the main body of the Sgr dwarf galaxy.
We selected the best GCs as those objects that have significant overdensities above the stellar background of the Sgr galaxy and that possess color-magnitude diagrams (CMDs) with well-defined red giant branches (RGBs) consistent with the distance and reddening of this galaxy. 
}
{
{We discover  eight new GC members of the Sgr galaxy, which adds up to 29 total GCs known in this dwarf galaxy. }
This total number of GCs shows that the Sgr dwarf galaxy hosts a rather rich GC system.
Most of the new GCs appear to be predominantly metal-rich and have low luminosity.
In addition, we identify {  ten} other GC candidates that are more uncertain and need more data for proper confirmation.
}

\titlerunning{Eight new globular clusters in the Sgr dwarf galaxy}
\authorrunning{Dante Minniti, et al.}

   \maketitle
   
\section{Introduction}
Globular cluster (GC) systems are useful to study the formation and evolution of galaxies. 
A number of recent theoretical models have been recently developed to study the GCs in their galactic context (e.g., Boylan-Kolchin 2017, Bose et al. 2018, Forbes \& Remus 2018, Pfeffer et al. 2018, Choski \& Gnedin 2019, Hughes et al. 2019, El Badry et al. 2019, Kruijssen et al. 2019, 2020, Burkert \& Forbes 2020), placing constraints on past merging events (e.g., kind of merger, galaxy masses, redshift of the accretion event, number of accreted GCs).
Observationally, however, there are only a handful of very nearby galaxies (within a 100 kpc volume) that contain GCs where we can measure their properties down to the faintest possible members: the Milky Way (MW), the Large Magellanic Cloud (LMC), the Small Magellanic Cloud (SMC), and the Fornax and Sagittarius (Sgr) dwarf galaxies, in order to compare them with the GC systems of other prototypical Local Group members, such as M31 and M33. 
As often occurs in Astrophysics, the knowledge of more distant systems in the Universe rests heavily on these few local standards.

In spite of their brightness, GCs can be hard to detect especially in some cases where the field stellar background density is very high and/or the reddening is high and inhomogeneous.
In the Galactic bulge, these effects are combined resulting in the incompleteness of the GC sample. Large numbers of cluster candidates have been discovered, many of which still need to be confirmed (e.g., Borissova et al. 2014, Barba et al. 2015, Minniti et al. 2017a, Palma et al. 2019).
In the LMC, even though the stellar density is high,  the reddening is not so extreme, and the foreground halo contamination is not so severe.
In consequence, LMC clusters can be more easily found and measured. 
Thousands of star clusters have been discovered in the LMC, exhibiting a wide range of ages, including classical old and very metal-poor GCs as well as younger and more metal-rich massive clusters 
(e.g., Searle et al. 1980, Da Costa 1991, Bica et al. 1996).
The GC system of Sgr, another one of those known galaxies within 100 kpc volume, is less studied, probably because of the relative recent discovery of this galaxy (Ibata et al. 1994). 
Adding the GC system of the Sgr galaxy to the local standards would be an important step.

The primary goal of this work is to find and characterize GCs in the Sgr dwarf galaxy, located at $D=26.5$ kpc behind the Milky Way bulge (Ibata et al. 1995, Monaco et al. 2004, Vasiliev \& Belokurov 2020).
However, in the case of the Sgr dwarf galaxy, finding star clusters is difficult mostly because of the high bulge stellar foreground density. 
Very few star clusters were previously known to be associated with this galaxy (Majewski et al. 2003, Belokurov et al. 2006, Law \& Majewski 2010, Forbes \& Bridges 2010, Massari et al. 2019, Myeong et al. 2019, Vasiliev 2019, Antoja et al. 2020).
According to the recent study of Bellazzini et al. (2020), there were nine GCs likely known to be Sgr members, four of them being located in the main body of this galaxy (NGC~6715, Arp~2, Ter~7, and Ter~8), and the other five being located in the extended tails (Pal 12, Whiting 1, NGC5634, NGC4147, and NGC2419).

Fortunately, the advent of Gaia has now made it possible to
deal with the severe bulge contamination problem by providing accurate proper motions (PMs; The Gaia Collaboration, Helmi et al.  2018, Brown et al. 2021), which allow us to efficiently discriminate foreground bulge stars.
We are now able to make clean Sgr density maps in order to search for star clusters that should be located at the right distance and should also be comoving with this galaxy.

Our past bulge GC searches with the VISTA Variables in the Via Lactea survey (VVV -- Minniti et al. 2010)  have identified several new GCs that are actually located behind the bulge in the Sgr dwarf galaxy.
We found a dozen new Sgr GCs, bringing the total number of GCs in this galaxy to 21 (Minniti et al. 2021).
Motivated by this success, we carried out a new search specifically tailored to find GCs within the main body of the Sgr galaxy for this study.
We concentrated on the central regions of Sgr, under the premise that GCs are brought in by dynamical friction (e.g., 
Gnedin \& Ostriker 1997, Arca-Sedda \& Capuzzo-Dolcetta 2014).
Our improved selection allowed us to find {  eight} new Sgr clusters, significantly increasing the sample of known GCs in this galaxy again.
We present their positions, significance, extinctions, number of RR Lyrae , as well as their optical and near-infrared (IR) color-magnitude diagrams (CMDs).

\begin{figure}[h]
\centering
\includegraphics[width=9cm, height=7.2cm]{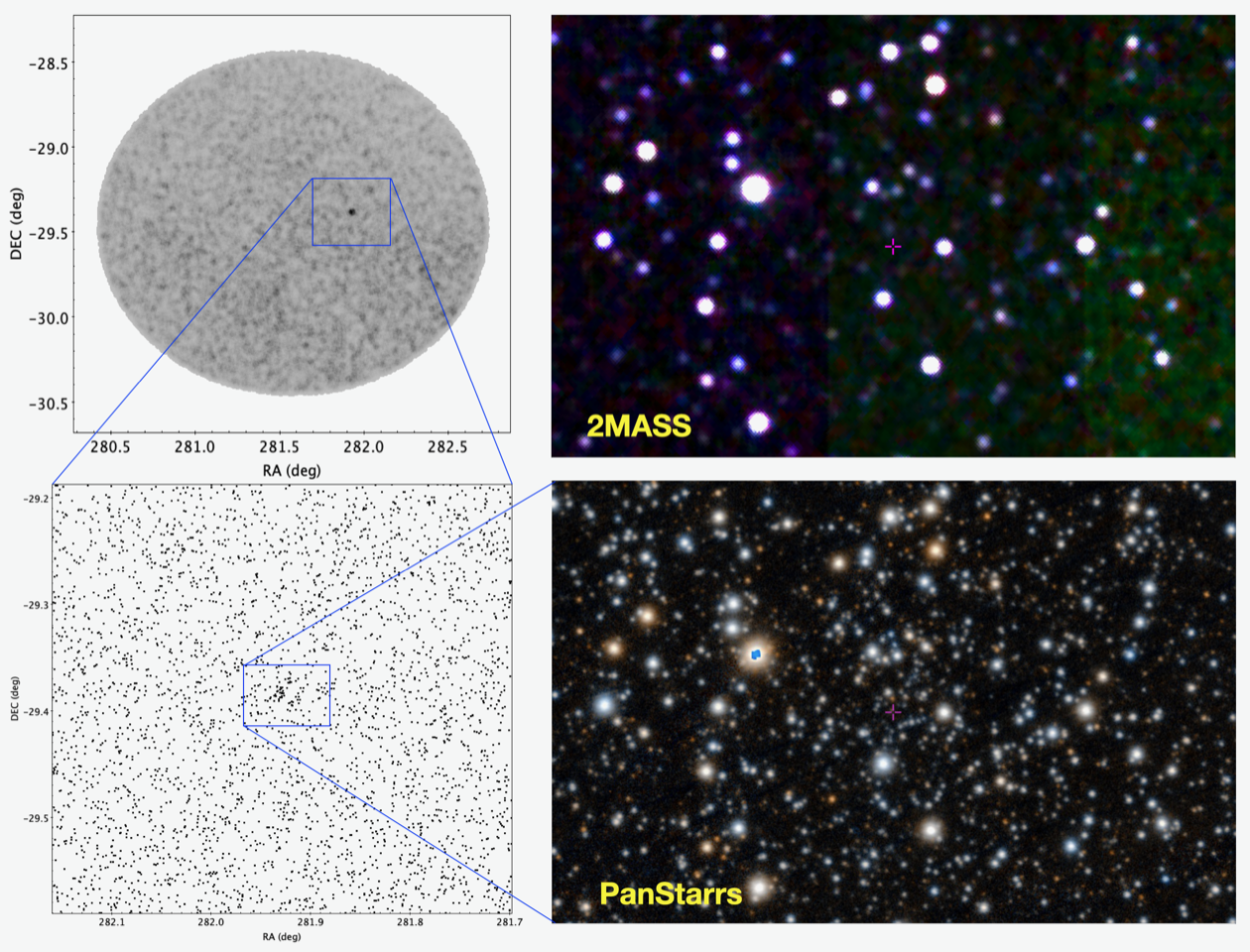} 
\caption{Maps for the new Sgr GC  Minni 332. 
Top left: Original Gaia EDR3 1 degree map containing all stars consistent with Sgr PMs.
Bottom left: Zoomed in field of about 15' as indicated.
Top right: Near-IR finding chart from 2MASS (Skrutskie et al. 2006). 
Bottom right: Optical color finding chart from Pan-Starrs (Magnier et al. 2021). 
Most of the bright stars in the optical and near-IR images are foreground Milky Way bulge stars. 
}
\end{figure}

\begin{figure}[h]
\centering
\includegraphics[width=9cm, height=6.2cm]{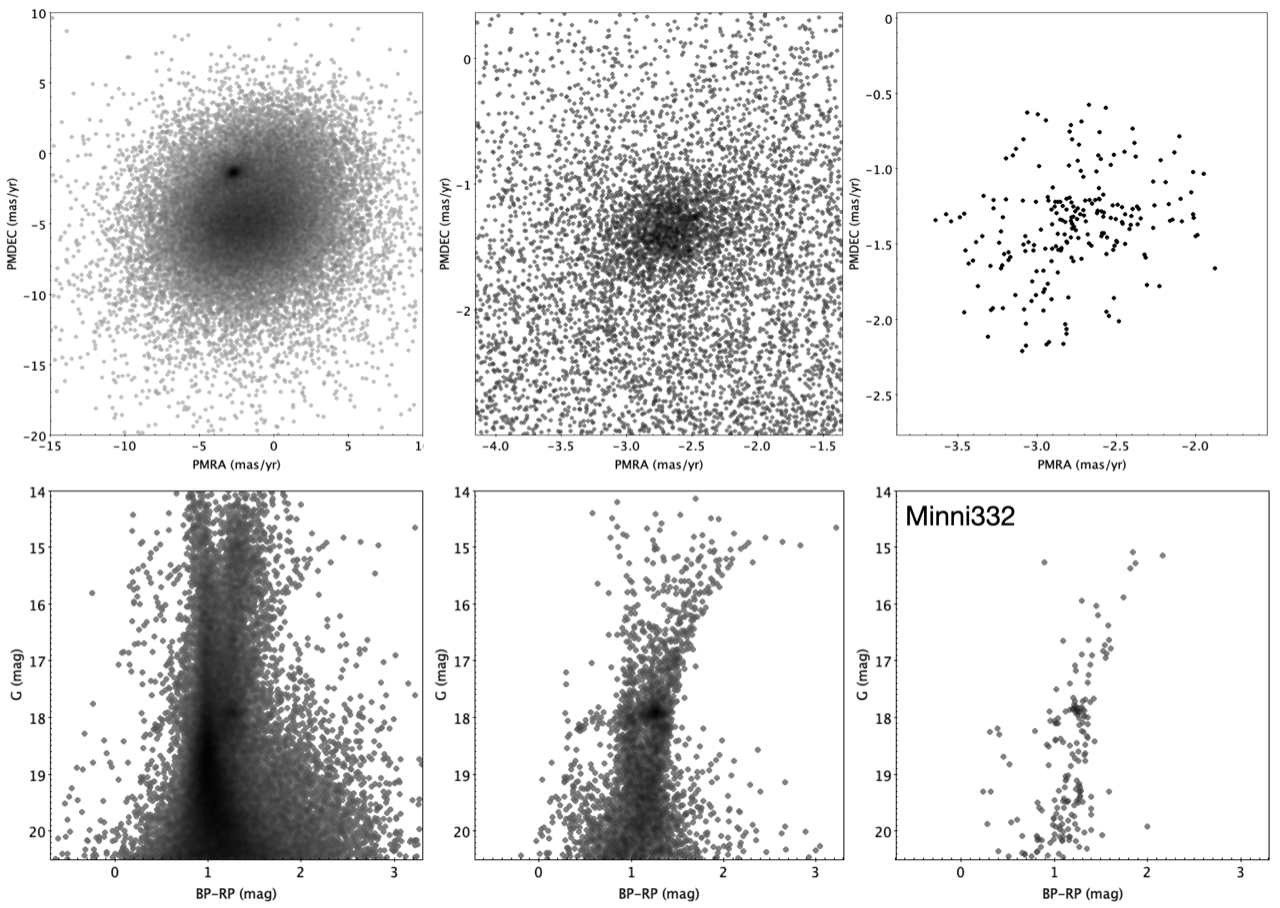} 
\caption{Vector PM diagrams (top panels) and CMDs (bottom panels) to illustrate the selection process, taking the GC Minni 332 as an example. 
Left panels: Original Gaia EDR3 diagrams containing all stars matched with near-IR VVVX photometry within a 1 degree field.
Middle panels: Subset of Sgr PM selected diagrams within a 1 degree field.
Right panels: { Final PM selected GC diagrams within 3 arcmin from the center of Minni 332 }.
}
\end{figure}

\section{Data and selection of new Sgr GCs} 

The GC selection strategy in Sgr needs to be slightly different than the one carried out in our bulge GC searches (Minniti et al. 2017a, 2021).  We pinpoint here its four main differences.
The first one is due to the effect of distance.
Because of their larger distances, the Sgr GCs should be roughly three times smaller than the bulge GCs and twice as large as the GCs of the Magellanic Clouds.
At the distance of Sgr, the scale is $2' \approx 15$ pc. Assuming a typical size as measured from their effective radii $r_{eff}$ for the MW GCs of 5 pc  (Harris 1996, edition 2010),  we expect Sgr GCs to have projected sizes that are roughly smaller than 1'.
The second difference involves the effect of extinction. Because Sgr is located at higher Galactic latitudes than the main bulge, extinction is not such a big problem. In the Galactic bulge, one has to consider the nonuniform reddening and the possibility that some of the cluster candidates may be mere regions with lower extinction than their surroundings (also known as dust windows).
The third difference is due to the effect of the line of sight depth. Assuming that the bulge has a radius of about 3.5 kpc, the bulge GCs can be located anywhere between 4.6 and 11.6 kpc from the Sun (adopting $R_{\odot}=8.18$ kpc, Gravity Collaboration 2019). However, the Sgr GCs would all be located at relatively the same distance. 
In fact, the difference in distance moduli between the near and far side of the bulge would be in excess of 2 magnitudes. On the other hand, allowing for a line of sight depth of 2 kpc within the Sgr dwarf galaxy, the difference in distance modulus from the near to the far side of this galaxy would be less than 0.2 mag. 
The last difference is due to the effect of the narrower range of PMs exhibited by the Sgr GCs, which should share the mean motion of this galaxy. The bulge GCs have a wide range of kinematics, but the Sgr galaxy is more distant and has lower velocity dispersion; therefore, its PMs are much more concentrated than the bulge.
This galaxy is also moving at a high speed toward the Galactic plane, and this characteristic mean motion actually helps to discriminate its associated objects.

For this new search, we used Gaia EDR3 optical data (The Gaia Collaboration, Brown et al. 2021) in combination with near-IR point spread function (PSF) photometry from the VVVX survey (Alonso et al. 2021, in preparation).
For two clusters located outside the VVVX footprint, we used the PSF photometry from McDonald et al. (2013).
The data provenance is indicated in Table 1, which is described in the Appendix.

Due to the  distance of the Sgr galaxy, it seems reasonable to assume that all of the reddening is located in the foreground.
At these relatively high southern Galactic latitudes ($b<-10^{\circ}$), the extinction seems to be fairly uniform. 
In fact, all the cluster candidates span a narrow range of extinctions and reddenings: $0.038<A_{Ks}<0.061$ mag and $0.053<E(J-Ks)<0.084$ mag, using the field reddening values from Schlafly \& Finkbeiner (2011).

\begin{figure*}[h]
\centering
\includegraphics[width=18.2cm, height=16.2cm]{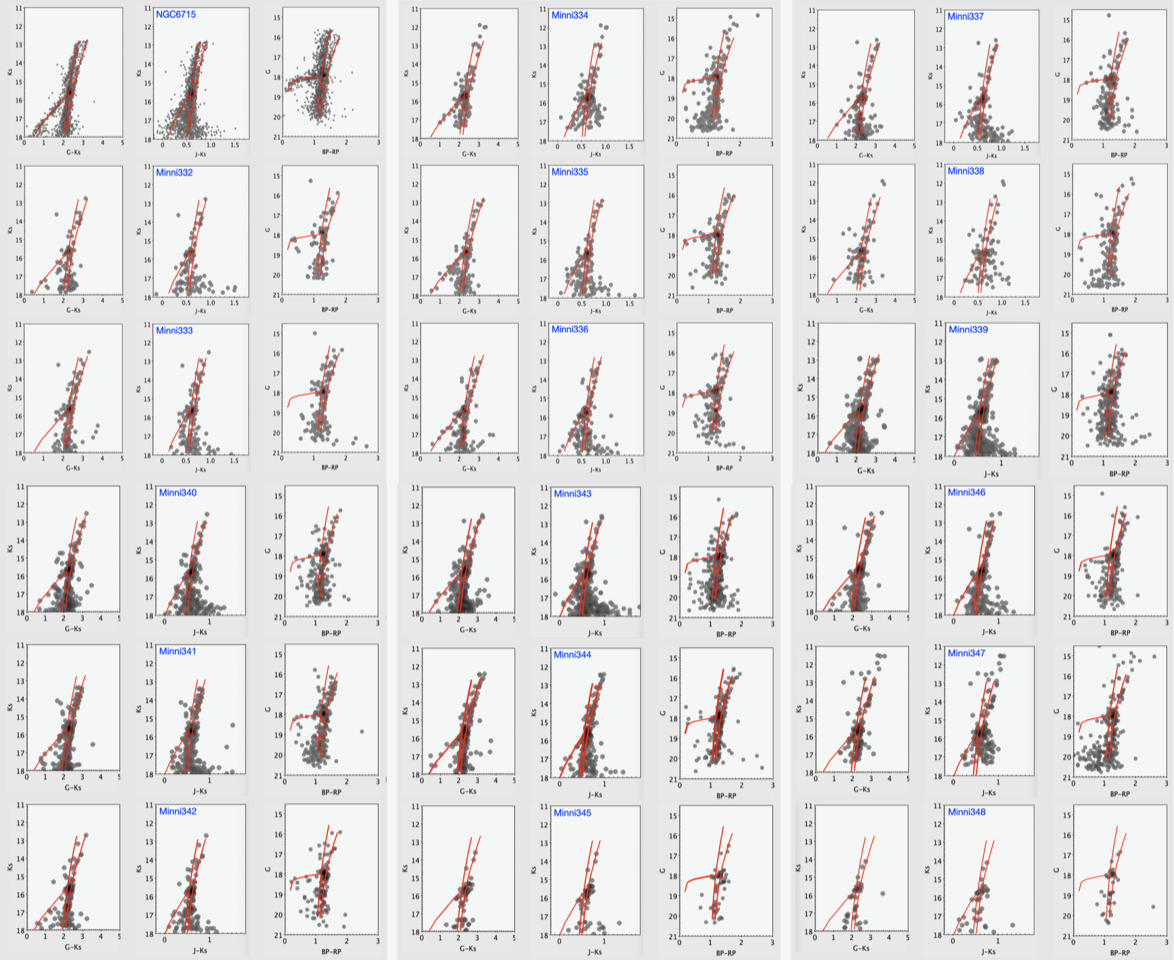} 
\caption{
Observed optical and
near-infrared color-magnitude diagrams for  the PM selected cluster members within 3 arcmin centered on the new candidate GCs. 
The mean RGB and horizontal branch (HB) ridge lines of the metal-poor GC NGC~6715 and the metal-rich Sgr field population are shown for comparison (left and right red solid lines, respectively).
}
\end{figure*}

The Gaia EDR3 vector PM diagram was used for member selection in order to efficiently discard foreground Milky Way disk and bulge stars along the line of sight toward the Sgr dwarf galaxy.
We chose Sgr members within $\pm 1$ mas/yr from the mean NGC 6715 PMs measured by Helmi et al. (2018) and Vasiliev \& Baumgardt (2021): $\mu_{\alpha}= -2.680 \pm 0.026$ mas/yr, $\mu_{\delta}= -1.387 \pm 0.025$ mas/yr. 
{  After trying different selections, we decided to adopt $\pm 1$ mas/yr as a compromise that takes the varying PM uncertainties at different magnitudes into account.}
{
Assuming isotropy, the mean Sgr RV dispersion is $\sigma= 15$ km/s for the region of interest (with a total range of $8 < \sigma <22$ km/s from Lokas et al. 2010). This implies a mean PM dispersion of $\sigma = 0.126$ mas/yr at $D=26.5$ kpc. Furthermore, this is to be compared with the GC velocity dispersions that are typically $\sigma \approx 5$ km/s, equivalent to 0.042 mas/yr at the distance of Sgr. Unfortunately, telling the difference between these dispersions is beyond the limit of Gaia measurement capabilities. 
}
We initially selected all the overdensities visually in the main body of Sgr, with approximate sizes between 1’ to 2’. A specific example (Minni 332) is shown in Figures 1 and 2.

\section{Results and discussion} 

All of the newly found cluster candidates are listed in Table 1.
The optical and near-IR CMDs (Figure 3) were then inspected to select the real GCs, by comparison with 
the CMDs of five known Sgr GCs in the region for which we have similar data (NGC~6715, Arp~2, Pal~12, Ter~7, and Ter~8).
{ NGC~6715 is also the largest of these GCs, with a tidal radius of $r_t=10'$ and a mean metallicity of $[Fe/H]=-1.30 \pm 0.12$ dex (Harris 1996, edition 2010, Baumgardt \& Hilker 2018, Fernandez-Trincado et al. 2021).
We note that Bellazzini et al. (2008) argued that NGC~6715 coincides with the nucleus of Sgr, but it is kinematically distinguished from the nucleus.}

{ 
In order to estimate the statistical significance of these stellar overdensities, we applied two tests.
First we followed the procedure of Koposov et al. (2007), } 
as applied in Moni Bidin et al. (2011) and Minniti et al. (2011, 2017b) who detected new Galactic GCs in the presence of heavy background contamination.
We computed the statistical significance of the stellar overdensities from the number of stars detected in excess to the local background, whose random fluctuations are assumed to be Poissonian (Koposov et al. 2007). 
As a specific example, we matched Ns = 211 stellar sources within 3’ centered in the GC Minni 332 (Figure 2), while the nearby background field density was measured to be Nb = 136 sources for a similar area. The excess of 75 sources was compared with the statistical error on the background number counts  $\sqrt N_b$, revealing a cluster detection at the $6.4 \sigma$ level.

{  Second, we also considered the variation of the background, taking the number counts of sources included within many adjacent circles of 3 arcmin radius around a wider area from the cluster candidates and using the standard deviation of the distribution of these number counts as sigma to compute the signal to noise. There are six GCs that survive the $3\sigma$ detection with this new method, and three others that maintain their status of bad candidates, so that both methods agree in the classification for nine sources. On the other hand, there are five gains that changed from bad to good using the new method and four losses that changed from good to bad in the new method. 

This illustrates that while the statistical significance can be computed in different ways, there are fluctuations. But in spite of these fluctuations, the really good candidates survive and the really bad ones do not, while in between a number of undefined candidates also remain. We conservatively chose to keep as real GCs only those that survive both determinations with $>3\sigma$ (Minni 332, 341, 342, 344, 348, and 349), labeled in boldface in Table 1. We also consider Minni 335 and 343 as very good candidates based on their CMDs and RR Lyrae content, even though they barely missed the second statistical criterion.

The significance of both detections is listed in Table 1 as $S/N_1$ and $S/N_2$, respectively,
where we consider  all the candidate GCs with  detections larger than $3.0$ to be significant.
There are six GCs that fulfill this requisite, to which we added Minni 335 and 343.
The remaining ten candidates have detections that are less significant, and they are considered unconfirmed. 
}
The overdensity in the selected clusters at the level of $>3 \sigma$ means that it is highly unlikely that they are all background fluctuations. 
{  However, we point out the need for follow-up spectroscopic observations and better Gaia EDR4 PMs for all of the cluster candidates listed in Table 1 in order to definitely confirm these GCs. }

{  The ten } unconfirmed candidates may not only be mere background fluctuations, but also real GCs in various stages of dissolution. It is not surprising to find dissolving clusters deep in the potential well of galaxies such as Sgr, as the dynamical processes that contribute to the GC demise are maximized in these inner regions (e.g., Arca Sedda \& Capuzzo Dolcetta  2014). As argued by Minniti et al. (2017a) and Palma et al. (2019) for the case of the numerous recently discovered bulge GC candidates, these objects are worthy of following up with additional observations in order to confirm their true nature.

We also examined which clusters may have associated RR Lyrae. 
After discarding the foreground bulge RR Lyrae on the basis of their PMs and mean magnitudes, three of these new GCs (Minni 335, 343, and 348) 
have four, three, and four RR Lyrae within 3’, respectively, which represents a $\approx 3 \sigma$ excess above the measured background of RR Lyrae from the Sgr galaxy.
All of these clusters have been classified as significant detections (Table 1).

In addition, about half of the new GCs exhibit extended blue HBs. 
However, none of the new GCs appear to be significantly metal-poor.
Figure 3 shows that in general the new GCs are more metal-rich than NGC~6715 ($[Fe/H]=-1.3$ dex) because  of their redder RGBs .

There are also two candidates, Minni 334 and 347, that have extended bright giant branches, indicating that they are either younger objects (more likely), or that they are located significantly in front of the Sgr dwarf galaxy (less likely).
One of the new GCs, Minni 341, lies very close to the Sgr nucleus (NGC~6715), only about 15' away. 
This GC is difficult to study because of the high field stellar density, but its detection is significant and its CMD is consistent with that of a typical GC (see Figures 3 and A.1).
We note that most of the new clusters are located in the central regions of this galaxy, where this search was performed, as shown in Figure 4. 
If these clusters were brought there by dynamical friction, their structures may not be normal, as they may be dissolving.

Following Minniti et al. (2021), we also compared the new GCs with previously known Sgr GCs for which we have similar data.
This comparison reveals that the new GCs are fainter than the comparison clusters Arp 2 ($M_V=-5.29$ mag), Ter~7 ($M_V=-5.01$ mag), and Ter~8 ($M_V=-5.07$ mag).

Apparently, all the previously known Sgr clusters were located to the south of NGC~6715, at the center of this galaxy. Also, only half of the Sgr GCs were located within the main body of the galaxy.
The new discoveries challenge this global picture. Now most of the Sgr GCs are located within the main body of this galaxy, and with many of them located to the north of the Sgr center, where our search was performed (Figure 4). In order to complete the spatial distribution of the Sgr GC system, however, a thorough search of the southern part of this galaxy is warranted.

\begin{figure}[h]
\centering
\includegraphics[width=7.2cm, height=7.0cm]{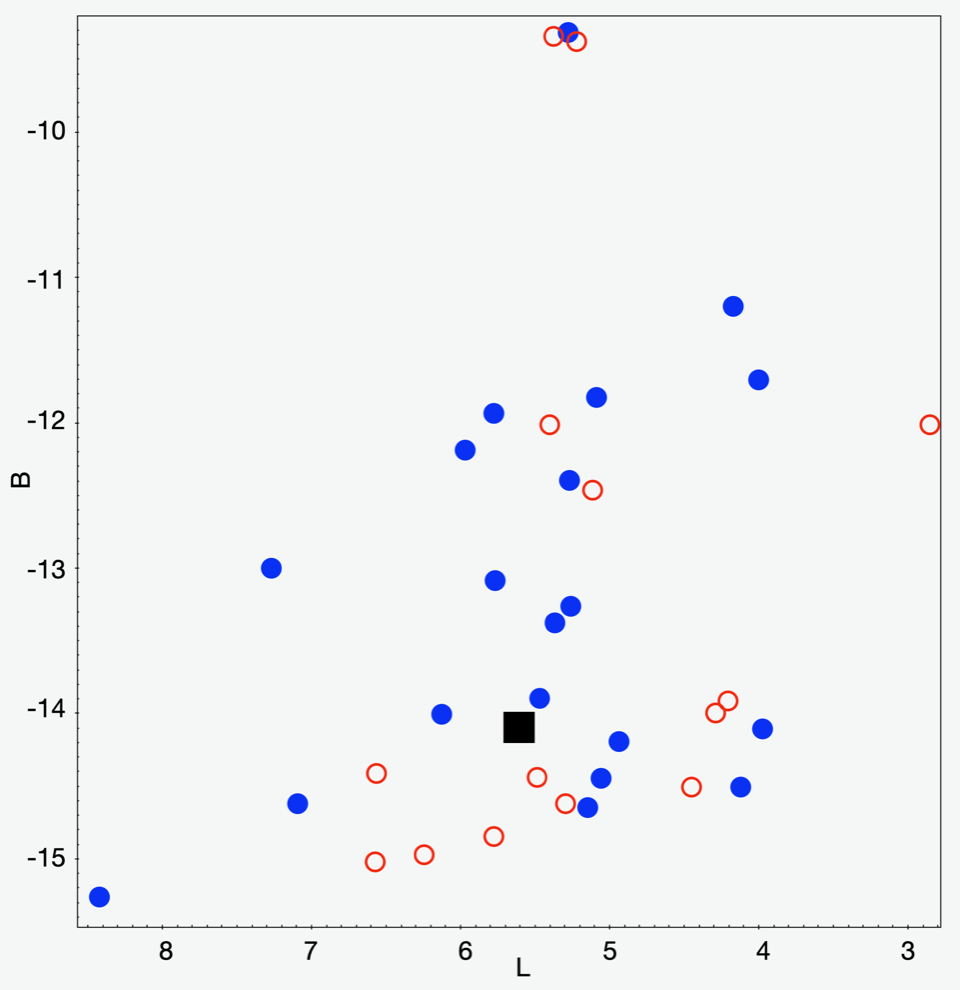} 
\caption{Galactic coordinate map showing the location of all of the new candidate GCs (red open circles) from Minniti et al. (2021) and this work.
The {  20 confirmed GCs} are highlighted with blue full circles. 
{ The position of the Sgr center that coincides with the location of the GC NGC 6715 is indicated with the large black square. }
We note that the southernmost part of the Sgr galaxy (at $b<-15^{\circ}$) still remains uncharted.
}
\end{figure}

\section{Conclusions}
{ We have carried out a new search for GCs 
in the main body of the Sgr dwarf galaxy} 
using the Gaia EDR3 optical data in combination with the near-IR data from the VVVX survey. 
This is, as far as we know, the first systematic optical and near-IR search for GCs within the main body of this galaxy.
For comparison, we used four known GCs that have compatible data: NGC~6715, Arp~2, Ter~7, and Ter~8.

In addition to the 12 new GCs found by Minniti et al. (2021), in this work we have identified {  eight} more GCs within the main body of the Sgr dwarf galaxy.
Even though the bulge field stars largely outnumber the cluster members, the exquisite Gaia EDR3 PMs allowed us to make clean optical and near-IR CMDs for the new GCs.
We also present their positions, extinctions, and detection significances.

After discarding the foreground bulge RR Lyrae variables stars, we found that three of the new Sgr GCs appeared to contain RR Lyrae.
Minni 335, 343, and 348 are confirmed to contain a significant overdensity ($\approx 3 \sigma$) of RR Lyrae above the nearby background.

We confirm that the GC system of the Sgr dwarf galaxy is richer than previously thought. The GC system of this galaxy now contains {  nearly 30} members.
However, from an observational standpoint, we conclude that the current census of Sgr GCs is still incomplete, as demonstrated by the continued discovery of faint GCs in the main body of this galaxy. We stress that a complete census of these objects has yet to be done in the southern portion of the Sgr dwarf galaxy, outside of the VVVX footprint.
Therefore, we predict the discovery of many more GCs with future facilities. 
In fact, these new GCs would also serve to train automatic detection algorithms to be applied to massive databases such as Gaia (The Gaia Collaboration, Brown et al. 2021), Pan-Starrs (Drlica-Wagner et al. 2020), and in the future the Vera Rubin Observatory  (also known as Large Synoptic Survey Telescope -- LSST Science Collaboration et al. 2009, Ivezic et al. 2008) in order to find even more missing GCs.

The next step is to measure the physical parameters for the new GCs: metallicities, ages, luminosities, and structural parameters (Garro et al. 2021). 
It would then be possible to compare the Sgr GC system with those of other well studied nearby galaxies such as the LMC.

Importantly, we are uncovering a new GC system of a very nearby galaxy that can be studied in detail. 
The Sgr galaxy is merging with the MW and its GCs may help to reveal the past history and also the future of this event, and of other GC systems of more distant galaxies as well.
The discovery of a populous GC system in the Sgr dwarf enables a variety of studies:
a comparison with the LMC and MW GC systems, which have widely different masses, potentials, and tidal fields;
a measurement of their orbital and structural parameters to explore their kinematical and dynamical evolution;
a measurement of their ages and metallicities to unveil their star formation history and chemical evolution
as well as to compare the age-metallicity relationship for these galaxies;
and a census of their variable star populations, including the RR Lyrae, to name a few.
\\

\begin{acknowledgements}
We gratefully acknowledge the use of data from the ESO Public Survey program IDs 179.B-2002 and 198.B-2004 taken with the VISTA telescope and data products from the Cambridge Astronomical Survey Unit. This work was developed in part at the Streams 21 meeting, virtually hosted at the Flatiron Institute.
D.M. and M.G. are supported by Fondecyt Regular 1170121 and by the BASAL Center for Astrophysics and Associated Technologies (CATA) through grant AFB 170002. 
J.A.-G. acknowledges support from Fondecyt Regular 1201490 and from ANID's Millennium Science Initiative ICN12\_009, awarded
to the Millennium Institute of Astrophysics (MAS). R.K.S. acknowledges support from CNPq/Brazil through project 305902/2019-9.
\end{acknowledgements}


{\bf References}
\\

Antoja, T., Ramos, P., Mateu, C., et al. 2020, A\&A, 635, L3

Arca-Sedda, M., \& Capuzzo-Dolcetta, R. 2014, ApJ, 785, 51

Barba, R., Roman-Lopes, A., Nilo Castellón, J. L., et al. 2015, A\&A, 581A, 120B

Baumgardt, H., \& Hilker, M., 2018, MNRAS, 478, 1520

{ Bellazzini, M., Ibata, R. A., Chapman, S. C., et al. 2008, AJ, 136, 1147B }

Bellazzini, M., Ibata, R., Malhan, K., et al. 2020, A\&A, 636, A107

Belokurov, V., Zucker, D. B., Evans, N. W., et al. 2006, ApJ, 642, L137

Bica, E., Clariá, J. J., Dottori, H., Santos, Jr., J. F. C., \& Piatti, A. E. 1996, ApJS, 102, 57

Borissova, J., Chené, A.-N., Ramírez Alegría, S., et al., 2014, A\&A, 569, A24

Bose, S., Ginsburg I. \& Loeb A., 2018, ApJ, 859, L13 

Boylan-Kolchin M., 2017, MNRAS, 472, 3120

Burkert, A., \& Forbes, D. 2020, AJ, 159, 56

Choski, N., \& Gnedin, O. Y. 2019, MNRAS, 488, 5409

Da Costa, G. S. 1991, in IAU Symposium 148, The Magellanic Clouds, eds. R. Haynes \& D. Milne, 183

Drlica-Wagner, A., Bechtol, K., Mau, S., et al. 2020, ApJ, in press (arXiv:1912.03302)

El Badry, K., Quataert, E., Weisz, D. R., et al. 2019, MNRAS, 482, 4528

{  Fernández-Trincado, J. G.,  Beers, T. C.,  Minniti, D., et al. 2021, A\&A, 648A, 70F }

Forbes D. A., \& Bridges T., 2010, MNRAS, 404, 1203

Forbes D. A. \& Remus R.-S., 2018, MNRAS, 479, 4760 

Gaia Collaboration, Brown, A. G. A., Vallenari, A., et al. 2018, A\&A, 616, A1 

{  Gaia Collaboration, Brown, A. G. A., Vallenari, A., et al. 2021, A\&A, 649A, 1G }

Gaia Collaboration, Helmi, A., van Leeuwen, F., et al. 2018, A\&A, 616, A12

{  Garro, E. R., Minniti, D., Gomez, M., et al. 2021, A\&A, submitted }

Gnedin, N.Y., \& Ostriker, J.P, 1997, ApJ, 486, 581G

Gravity Collaboration, Abuter, R., Amorim, A., et al. 2019,  A\&A, 625, L10

Harris, W. E., 1996, AJ, 112, 1487

Helmi, A., Babusiaux, C., Koppelman, H. H., et al. 2018, Nature, 563, 85

Hughes, M. E., Pfeffer, J., Martig, M., et al. 2019 MNRAS, 482, 2795

Ibata, R. A., Gilmore, G., \& Irwin, M. J. 1994, Nature, 370, 194

Ibata, R. A., Gilmore, G., \& Irwin, M. J. 1995, MNRAS, 277, 781I

Ivezic, Z., Tyson, J. A., Abel, B., et al. 2008, arXiv:0805.2366

Koposov, S.; de Jong, J. T. A.; Belokurov, V., et al. 2007, ApJ, 669, 337K

Kruijssen, J. M., Pfeffer, J. L., Reina-Campos, M., et al. 2019, MNRAS, 486, 3180

Kruijssen, J. M., Pfeffer, J. L., Chevance, M., et al. 2020, MNRAS, 498, 2472

Law, D., \& Majewski S. 2010, ApJ, 718, 1128

{ Lokas, E. L., Kazantzidis, S., Majewski, S. R., et al., 2010, ApJ, 725, 1516}

LSST Science Collaboration et al., 2009, preprint, arXiv:0912.0201

Magnier, E. A., Chambers, K. C., Flewelling, H. A., et al. 2021, ApJS, 251, 3M  

Majewski, S. R., Skrutskie, M. F., Weinberg, M. D., \& Ostheimer, J. C. 2003, ApJ, 599, 1082

Massari, D., Koppelman, H. H., \& Helmi, A. 2019, A\&A, 630, L4

McDonald, I., Zijlstra, A. A., Sloan, G. C., et al., 2013, MNRAS, 436, 413 

Minniti, D., Geisler, D., Alonso Garcia, J., et al., 2017a, ApJ, 849, L24

{ Minniti, D., Hempel, M., Toledo, I., et al. 2011, A\&A, 527, A81}

Minniti, D., Lucas, P. W., Emerson, J. P., et al. 2010, NewA, 15, 433 

{ Minniti, D., Palma, T., Dekany, I., et al. 2017b, ApJ, 838L, 14M}

Minniti, D., Ripepi, V., Fernández-Trincado, J. G., et al. 2021, A\&A, 647, L4M

Monaco, L., Ferraro, F. R., Bellazzini, M., \& Pancino, E. 2004, MNRAS, 353, 874M

{ Moni Bidin, C., Mauro, F., Geisler, D., et al. 2011, A\&A, 535, A33}

Myeong, G.C., Vasiliev, E., Lorio, G., et al. 2019, MNRAS, 488, 1235M

Palma T., Minniti, D., Alonso-Garcia, J., et al., 2019, MNRAS, 487, 3140

Pfeffer, J., Kruijssen J. M. D., Crain R. A., \& Bastian N., 2018, MNRAS, 475, 4309

Saito, R. K., Hempel, M., Minniti, D., et al. 2012, A\&A, 537, A107

Schlafly, E. F., \& Finkbeiner, D. P., 2011, ApJ, 737, 103

Searle, L., Wilkinson, A., \& Bagnuolo, W. G. 1980, ApJ, 239, 803

Skrutskie, M. F., Cutri, R. M., Stiening, R., et al. 2006, AJ, 131, 1163

Vasiliev, E., 2019, MNRAS, 484, 2832

Vasiliev, E.,  \& Baumgardt, H., 2021, MNRAS in press (arXiv:2102.09568)

Vasiliev, E.,  \& Belokurov, V., 2020, MNRAS 497, 4162

\begin{table}
\onecolumn
\centering 
\caption{New GC candidates in the Sgr galaxy}
\begin{tabular}{lcccrrllll}
\hline\hline
ID             &RA      & DEC  & $A_{Ks}$ & N$_3$  & N$_{10}$  & $S/N_1$ & $S/N_2$ &  Survey  \\          
                & (J2000)& (J2000)&[mag]   &         &          &  \\
\hline
{\bf  Minni 332}     &18 47 43.2 &-29 23 06  &0.045  & 1  &7 &6.4  &5.2 &VVVX \\
Minni 333     &18 58 07.2 &-29 45 36  &0.051         & 1  &8 &2.2  &6.7 &VVVX \\
Minni 334     &18 59 57.6 &-30 16 12  &0.038         & 1  &9 &2.6  &1.7 &VVVX \\
{\bf  Minni 335}    &18 47 16.8 &-30 15 36  &0.054  & 4 &12 &5.5  &2.2 &VVVX  \\
Minni 336     &18 51 48.0 &-31 39 36  &0.049        & 0  &2  &2.7  &3.4 &VVVX \\
Minni 337     &18 52 16.8 &-31 37 12  &0.050        & 0  &2  &1.8  &2.3 &VVVX \\
Minni 338     &18 54 52.8 &-31 40 48  &0.047      & 0  &1  &3.7  &2.0 &VVVX  \\
Minni 339   &18 45 55.2  &-29 49 12  &0.042         & 0 &8   &2.7  &4.9 &VVVX  \\
Minni 340   &18 50 55.2  &-30 27 36  &0.046       & 2 &22 &3.4  &1.9 &VVVX \\
{\bf   Minni 341}   &18 53 57.6  &-30 31 48  &0.043  & 2 &11 &4.9  &4.8 &VVVX \\
{\bf   Minni 342}   &18 54 19.2 &-31 07 12  &0.042   & 1 &5   &4.6  &3.9 &VVVX \\
{\bf   Minni 343}   &18 55 36.0 &-31 06 36  &0.045   & 3 &13  &4.6  &2.4 &VVVX \\
{\bf   Minni 344}   &18 56 38.4  &-31 06 36  &0.045  & 1 &15 &3.2  &3.8 &VVVX \\
Minni 345   &18 56 45.6  &-30 58 12  &0.044          & 2 &11 &1.6  &12.5&VVVX \\
Minni 346   &18 56 21.6 &-30 43 48  &0.044          & 2 &12  &1.4  &0.1 &VVVX  \\
Minni 347   &18 58 36.0  &-30 37 48  &0.042         & 1 &15 &2.0  &3.4 &MD \\
{\bf   Minni 348}  &18 59 57.6  &-29 22 12  &0.046  & 4 &14  &4.4  &8.1 &VVVX \\
{\bf   Minni 349}  &19 04 57.6  &-28 26 24  &0.061  & 0 &3   &3.5  &3.3 &MD \\
\hline\hline
\end{tabular}
\label{rrlpat}
\end{table}

\begin{appendix}
\section{Sgr GCs data}

Table A1 below lists the data for all the GC candidates analyzed. 
We give the IDs, 
positions in equatorial coordinates (J2000), 
near-IR extinctions $A_{Ks}$,
number of RR Lyrae within 3' and 10' from the cluster centers (N$_3$ and N$_{10}$), 
significance of the overdensity, 
and data provenance VVVX or MD (McDonald et al. 2013). 
Confirmed GCs ($\sigma > 3.0$) are in boldface.

Figure 5 below shows 
the optical  CMDs for the best clusters listed in Table 1 compared with their respective field CMDs (with only PM-selected Sgr members and equal areas being considered). 
We looked for differences that can be appreciated with respect to the Sgr field stars, including the following, for example: 
the width of the RGB, which is larger in the Sgr dwarf than in the typical GCs; the presence of young and intermediate age stars in the Sgr body, which are absent in old GCs, and in some cases of metal-poor GCs; and the presence of an extended HB. 
In fact the CMDs for the well populated GCs are different than the typical Sgr population, showing instead similar features to the known Sgr GCs used for comparison (NGC~6715, Arp~2, Pal~12, Ter~7, and Ter~8), with CMDs shown by Minniti et al. (2021).

\begin{figure*}[h]
\centering
\includegraphics[width=12.2cm, height=14.2cm]{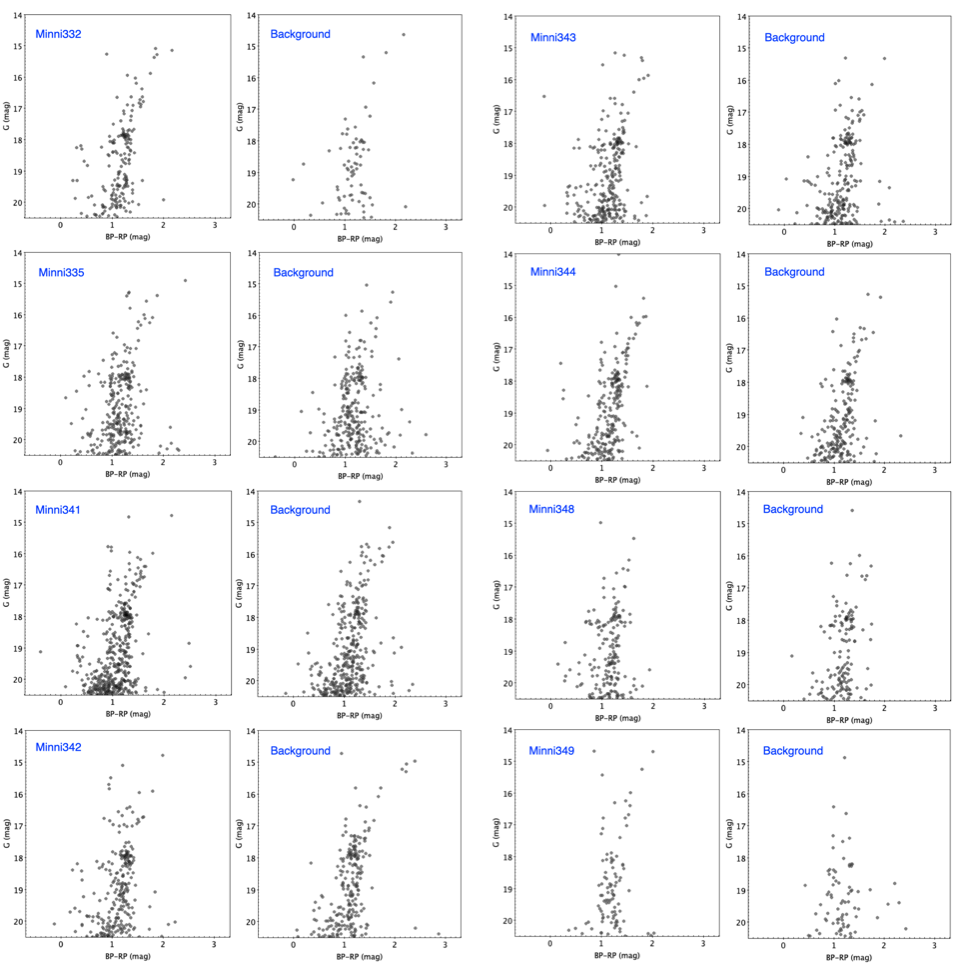} 
\caption{
{ GC CMDs from Gaia EDR3 photometry compared with their respective field CMDs.}
}
\end{figure*}

\end{appendix}

\end{document}